\documentclass[onecolumn]{aa}
\usepackage{graphicx}

\begin{document}

%   \thesaurus{11     % A&A Section 11: Galaxies
%              (02.18.15; % Radiation mechanisms: non-thermal,
%              02.13.1;  % Magnetic fields,
%               02.16.1;  % Plasmas,
%               11.10.1)} % Galaxies: jets.
%
\title{Relativistic parsec--scale jets: \uppercase{i}. Particle acceleration}

\author
{A.R. Beresnyak\inst{1}
        \and
        Ya.N. Istomin\inst{1}
        \and
        V.I. Pariev\inst{1}\inst{2}}

        \offprints{Ya.N. Istomin}

        \institute{
P.~N.~Lebedev Physical Institute,
Leninsky Prospect 53, Moscow 117924, Russia \\
email: istomin@td.lpi.ac.ru
    \and
Department of Physics and Astronomy, University of Rochester,
Rochester, NY 14627, USA email: vpariev@pas.rochester.edu }

        \date{Received  ; accepted  }

\titlerunning{Relativistic parsec--scale jets: \uppercase{i}}
\authorrunning{A.R. Beresnyak et al.}

%\maketitle

\abstract{
We develop a theory of particle acceleration inside relativistic
rotating electron-positron force--free jets with spiral magnetic
fields. We considered perturbation of the stationary magnetic
field structure and found that acceleration takes place in the
regions where the Alfv\'en resonant condition with the eigenmodes
in the jet is fulfilled, i.e. where the local Alfv\'en speed is
equal to the phase speed of an eigenmode. The acceleration
mechanism is regular acceleration by the electric field of the
electromagnetic wave, which is the eigenmode of the force--free
cylindrical jet: particles drift out of the region of the large
wave amplitude near the Alfv\'en resonant surface and gain energy.
Acceleration in the strong electric field near the Alfv\'en resonance
and synchrotron losses combine to form a power-law energy spectrum
of ultra-relativistic electrons and positrons with index between 2
and 3 depending upon the initial energy of the injected particles.
The power law distribution ranges from $\sim 10\mbox{MeV}$ to
$\sim 10^3\mbox{MeV}$.

      }

 % \end{abstract}

\maketitle

\section{Introduction}
The nature of the Active Galactic Nuclei (AGN) is still unclear.
The most common point of view is that there is a supermassive
black hole in the centre of the active galaxy with a mass
approximately $10^8$ to $10^9 \, M_{\sun}$ (Rees, 1984). Accretion
onto the black hole drives powerful turbulent processes in
accretion discs which effectively heat the plasma and generate a
magnetic field of the order of $10^4\,\mbox{G}$. Ejection of
plasma from discs with frozen magnetic field and radiation from
the disc lead to the formation of collimated streams which
transfer energy to large distances, up to $10^5$ pc. In addition, the
processes of generation of electron--positron plasma may occur in
the magnetospheres of black holes as well as in the vicinity of
accretion discs. Particles are created in collisions of
high--energy photons produced by inverse Compton scattering of
ultraviolet photons emitted from the disc with fast particles
accelerated by magnetic reconnection within the disc or  within
the magnetosphere of the rotating black hole, where the equilibrium
charge density changes its sign (Beskin et al. 1992;
Hirotani~\& Okamoto, 1998). Electron--positron plasma forms
pinched streams (jets) which have radii of the order of one
parsec. The electromagnetic energy density in these jets greatly
exceeds the total energy density of electrons and positrons,
including the rest energy. The angular momentum from the black hole and
the accretion disc is transmitted to the jet via magnetic
coupling, where a strong radial electric field $E_r$ arises. Drift
in this field and in the longitudinal magnetic field, $B_z$,
frozen into plasma leads to differential rotation. The jet
transfers an electric current producing an azimuthal magnetic
field, $B_\phi$, and magnetic field lines twist into spirals.
The energy in the jet is transferred by the Pointing vector which is
proportional to the product of the radial electric field and the azimuthal
magnetic field. Therefore, there is no problem of energy transfer
along the jet to large distances compared to the case where the
main energy is in the hydrodynamic motion.

The amazing facts about the jets from AGN are their high
collimation (the ratio of the length to the diameter is a few
orders of magnitude), the stability, and apparent superluminal
velocities of distinct knots (Begelman et al. 1984).
The hydrodynamic and MHD stability of jets was investigated in
many studies (Blandford \& Pringle, 1976; Torricelli--Ciamponi \&
Petrini, 1990; Appl \& Camenzind, 1992). The stability of the
electron--positron jets, rotating and moving with relativistic
speed and surrounded by a dense medium was studied by Istomin \&
Pariev~(1994) and Istomin \& Pariev~(1996). Such jets were shown
to be stable with respect to axisymmetric as well as spiral
perturbations. The physical reason for the stability is the shear
of the magnetic field, which in the case of low hydrodynamic
pressure stabilizes small oscillations. The density of the media
surrounding the jet, $\rho$, should satisfy the condition $\rho
\gg B^2/(4\pi c^2)$ for the inertia of surrounding gas to
stabilize the relativistic jet. On parsec scales, a typical value
of $B \sim 10^{-2}\,\mbox{G}$ gives a proton density $n \gg
0.05\,\mbox{cm}^{-3}$. The characteristic density of gas in
galactic nuclei is $n \sim 10\,\mbox{cm}^{-3}$. Therefore, the
approximation of stationary jet walls (Istomin~\& Pariev, 1994;
1996) is well justified. A relativistic force--free jet with
infinite radial extent or with vacuum outside can develop weak
instabilities (Lyubarskii, 1999; Li, 2000). In a stable jet,
perturbations do not increase with time
($\mbox{Im}\,\omega\equiv0$) or have a small decay rate
($\mbox{Im}\,\omega\approx10^{-2} \mbox{Re}\,\omega$) because of
the resonance with Alfv\'en waves $\omega^\prime=k_\|^\prime c$.
When the specific energy density and pressure of the plasma are
much less than the energy density of the magnetic field, the
Alfv\'en velocity is equal to the speed of light in vacuum $c$
($\omega^\prime$ and $k^\prime$ are the frequency and the wave
vector in the plasma rest frame). The resonance condition is
fulfilled on the specific magnetic surface and, in the case of
cylindrical jet, at a definite distance from the axis. In the
vicinity of that surface the magnetic and electric fields of the
wave are large. Particles in the jet are accelerated by the
electric field there, absorbing the energy of the perturbation.
Thus, the stability of the jet is directly related to the
production of energetic particles in the jet. This resolves the
problem of {\it in situ} acceleration of electrons and positrons
producing synchrotron emission from knots observed along jets. In
fact, as  is known (see Begelman at al., 1984), the energetic
particles accelerated in the central region cannot survive far
along the jet; they must lose their transverse momentum by
synchrotron radiation already at the basement of the jet. The time
for an electron to lose its perpendicular momentum due to
synchrotron radiation is given by
 $$ \tau_{\rm s}=\frac34\frac{c\gamma_\|}{r_{\rm e}\omega_{B0}^2}. $$
Here $\gamma_\|$ is the longitudinal Lorentz--factor of the
particle, $\gamma_\|=(1-v_\|^2/c^2)^{-1/2}$; $r_{\rm e}$ is the
classical radius of the electron $r_{\rm e}=e^2/m_{\rm e} c^2$;
$\omega_{B0}=eB/m_{\rm e} c$ is the electron cyclotron frequency.
Given  $\gamma_\|=10$, $B=10^2\,\mbox{G}$, $\tau_{\rm s}$ is
approximately $3\cdot10^5\, \mbox{sec}$, whereas the typical time
of the flyby through the region of strong fields $\tau_{\rm
f}\approx \ell/c\approx 10^6\, \mbox{sec}$. Here $\ell$ is the
diameter of the jet, $\ell\approx 3\cdot10^{16}\,\mbox{cm}$.
Qualitatively the possibility of particle acceleration by
electromagnetic disturbances propagating along a jet was proposed
by Bisnovatyi--Kogan \& Lovelace (1995). However, the authors of
this work needed an ad hoc vacuum gap between conducting jet and
surrounding medium, where the acceleration is possible due to the
presence of the component of electric field parallel to the
magnetic field.

In this paper we assume that the energy transferred along the jet
is mainly the electromagnetic energy which can be transmitted for
long distances along the jets as if along wires. In this case it
is very possible that part of that energy is in the waves
propagating along the jet, which are eigenmodes of a cylindrical
beam. The source of the wave motion consists of non-stationary processes in
the magnetospheres of the black hole and the accretion disc. Short
time variability on scales from days to months is actually
observed in AGN~(Mushotzky et al. 1993; Witzel et al.,
1993). Such a picture also gives a natural explanation of the
superluminal motions but requires neither special orientation of
the jet nor a relativistic jet with high speed
($\beta=u/c>\sqrt2/2$). As was shown in papers by Istomin~\&
Pariev~(1994) and Istomin~\& Pariev (1996) there exist so called
standing eigenmodes in jets ($v_{\rm group}\equiv0$) which do not
propagate with finite velocity along the jet but are only subject
to diffuse spreading and, therefore, their amplitude is maximal.
The phase velocity of these modes is greater than $c$. Wave crests
move along the jet  with superluminal velocity causing the
acceleration of the particles on the Alfv\'en surface. These
regions might be the observed bright knots with typical sizes of
the order of the wavelength of the standing wave, which is about
the radius of the jet.

There is some observational evidence in the case of 3C273, M87,
and 3C345 that their jets are electron-positron rather than
electron-proton (Morrison \& Sadun, 1992; Reynolds et al., 1996;
Hirotani et al., 2000). The recent observation of circular
polarization of radio emission from 3C279, 3C273, 3C84 and
PKS0528+134 (Wardle et al., 1998 ) supports the same conclusion.
However, there are also indications that jets in Optically
Violently Variable quasars cannot consist solely of e$^{+}$e$^{-}$
pairs because they would produce much larger soft X-ray
luminosities than are observed. On the other hand, models with
jets consisting solely of proton-electron plasma are excluded,
since they predict much weaker nonthermal X-radiation than
observed in Optically Violently Variable quasars (Sikora~\&
Madejski, 2000).

We consider the acceleration processes near the Alfv\'en resonant
surface and calculate the spectrum of accelerated particles. In an
accompanying paper (Pariev et al. 2002), we
calculate the polarization of synchrotron radiation generated by
these particles.

\section{Acceleration of particles}

The equations of motion near the Alfv\'en surface are rather
complicated. Nevertheless, we can use drift equations~(Sivukhin,
1965) because the Larmor radius of relativistic electrons and
positrons $r_{\rm c}= {\cal E}/eB\approx10^6\mbox{cm}$ (if we take
$B\approx 10^{-2}$ and $\gamma\approx 10$) is much smaller than
radius of the jet and the width of the resonant surface $r_0$ (see
expression (28)):
\def\e{{\cal E}}
$$ \frac{d{\bf r}}{dt}=v_\|\frac{{\bf B}}B+\frac c{B^2}[{\bf E\times B}]+
                   \frac{\e v_\|^2}{ecB^4}[{\bf B\times(B\nabla)B}]+
                   \frac{\e v_\perp^2}{2ecB^3}[{\bf B\times\nabla}B];$$
$$
 \left(\frac{d\e}{dt}\right)_{\rm e.m.}=e{\bf E}\frac{d{\bf r}}{dt}+
\frac{\e v_\perp^2}{2c^2B}\frac{\partial B}{\partial t}.\eqno (1)
$$
Here the electric field, ${\bf E}$, is assumed to be small
compared with the magnetic field, ${\bf B}$, and $\e$ is the
energy of the particle. The index $\mbox{e.m.}$ means the energy
change due to the action of the electromagnetic fields, ${\bf E}$
and ${\bf B}$ on the particle. We did not write the third equation
here, the conservation of the adiabatic invariant
$J_{\perp}=p_{\perp}^2/B$, which is usually used as one of the
drift equations. The adiabatic invariant is not conserved because
of synchrotron losses. But the characteristic time of the change
of the adiabatic invariant is much larger than the period of the
cyclotron rotation and we can use the drift approximation of the
motion of the particle as a result of the separation of the motion
into fast cyclotron rotation and slow motion of the guiding
centre with subsequent averaging over the fast rotation. The
conditions for using the drift approximation are a small change of
the cyclotron frequency $\omega_B$ during cyclotron period
$2\pi/\omega_B$ and also a small change of the particle cyclotron
radius on the scale of the cyclotron radius (Alfv\'en~\&
F\"althammar, 1963):
$$
\frac{2\pi}{\omega_B}\frac{\partial \omega_B}{\partial t} \ll
\omega_B; \quad r_B |\nabla r_B| \ll r_B. \eqno (1a)
$$
A rigorous derivation of equations~(1) under conditions~(1a) was
given by Sivukhin~(1965) and Morozov~\& Solov'ev (1966). Let us
demonstrate that the second equation~(1) does not require the
conservation of the adiabatic invariant $J_{\perp}$. The particle
velocity is split into two parts ${\bf v} = d{\bf r}/dt + {\bf
v}_{\rm c}$. Here $d{\bf r}/dt$ is the velocity of the guiding
centre and ${\bf v}_{\rm c}$ is the velocity of cyclotron
rotation, $v_{\rm c} = v_\perp$.
$$
\left(\frac{d\e}{dt}\right)_{\rm e.m.} = e{\bf v}{\bf E} =
e\frac{d{\bf r}}{dt}{\bf E} + e{\bf v}_{\rm c}{\bf E} =
e\frac{d{\bf r}}{dt}{\bf E} + \frac{e\omega_B}{2\pi}\oint{\bf E}d{\bf l} = \\
e\frac{d{\bf r}}{dt}{\bf E} + \frac{e\omega_B}{2\pi}\int \>
\mbox{curl}\, {\bf E}\,d{\bf s}.
$$
Using the Maxwell equation, $\mbox{curl}\, {\bf E} =
-(\partial{\bf B}/\partial t)/c$, and the relation ${\bf
s}=-\pi(v_\perp/\omega_B)^2{\bf B}/B$ we obtain the second
equation of the drift approximation~(1). The first equation of
system~(1) is the equation of the motion of the guiding centre
$d{\bf r}/dt$ and contains the motion along the magnetic field
line, the electric drift, gradient and centrifugal drifts. All
drifts and the equation of energy change are independent of the
conservation of the adiabatic invariant $J_{\perp}$. The value of
$J_{\perp}$ is conserved in the slowly varying electromagnetic
fields ${\bf B}({\bf r},t)$, ${\bf E}({\bf r},t)$, but is not
conserved under the action of additional forces including
radiative forces. The small change of the adiabatic invariant
during the period of the cyclotron rotation does not violate the
condition of applicability of the drift approximation~(1a).

The electromagnetic fields are equal to the sum of the stationary
fields (without subscript) and the wave fields (subscript 1). For
a cylindrical jet the stationary configuration of fields is
(Istomin \& Pariev, 1994) (c=1):
\def\omr{\Omega^{\rm F} r}
$$ {\bf B}=B_z{\bf e}_z+B_\phi{\bf e}_\phi;\quad B_\phi=\omr B_z;$$
$$ {\bf u}=K{\bf B}+\omr{\bf e}_\phi;\eqno (2) $$
$$ {\bf E}=-\omr[{\bf e}_\phi{\times\bf B}], $$
 where $r,\phi,z$ are cylindrical coordinates. In components
$$ {\bf B}=(0,\omr,1)B_z,$$
$$ {\bf E}=(-\omr,0,0)B_z.$$
Here $\Omega^{\rm F}$ and $K$ are functions of $r$, and $B_z$ does
not depend on $r$. The stationary electric field, which has an
absolute value of $\omr B_z$, is not small compared to the
magnetic field when $\omr\sim 1$; therefore we must consider our
equations in the frame moving with the plasma where ${\bf E}\equiv 0$.
There are many such reference frames because there is an arbitrary
parameter $K(r)$ in the expression for the plasma velocity which
determines the radial profile of longitudinal velocity. $K(r)B_z$
is the velocity along the magnetic field which is not related to
the rotation. We choose, however, the velocity ${\bf u}$ which
minimizes the kinetic energy of the plasma in the stationary
reference frame.
$$ {\bf u}=\frac{(0,\omr,-(\omr)^2)}{1+(\omr)^2}.\eqno (3)$$
It is probable that the plasma moves with that velocity in real
jets (Frolov~\& Novikov 1998). THE fields in the wave calculated in
terms of the Lagrangian radial displacement $\xi$ are (Istomin~\&
Pariev 1996) (here we drop out the phase coefficient $\exp
(-i\omega t+ikz+im\phi)$):
$$ B_{r1}=iB_zF\xi;$$
$$ B_{\phi 1}=-B_z(\omr\frac{d\xi}{dr}+\xi\frac d{dr}(\omr)+\frac k S D);$$
$$ B_{z1}=B_z(-\frac{d\xi}{dr}-\frac\xi r+\frac m{rS} D);\eqno (4)$$
$$ E_{r1}=B_z(\omr\frac{d\xi}{dr}+\xi\frac d{dr}(\omr)-\frac\omega S D);$$
$$ E_{\phi 1}=-iB_z(\omega-m\Omega^{\rm F})\xi;$$
$$ E_{z1}=iB_z\omr(\omega+k)\xi;$$
where
$$ D=r\frac{d\xi}{dr}\left(\Omega^{\rm F}(\omega+k)-\frac m{r^2}\right)-
\xi\left(\Omega^{\rm F}(\omega+k)+\frac m{r^2}\right),$$
$$ F=k+m\Omega^{\rm F},$$
$$ S=\omega^2-k^2-m^2/r^2\mbox{.} $$
For Lagrangian displacement $\xi$ and dimensionless pressure
disturbance $p_*=4\pi P_1/B_z^2$ we have a system of differential
equations obtained by Istomin \& Pariev~(1996):
\def\fracb#1#2{\frac{\displaystyle #1}{\displaystyle #2}}
$$ \left\{\begin{array}{rcl}
     A\fracb1r\fracb d{dr}(r\xi)&=&C_1\xi-C_2p_*,\\
        A\fracb{dp_*}{dr}&=&C_3\xi-C_1p_*.\\
          \end{array}  \right.\eqno (5)
$$
Here
$$ C_1=\frac2r\left(m\Omega^{\rm F}-(\omr)^2(\omega+k)\right);$$
$$ C_2=-\frac{\omega^2-k^2-m^2/r^2}{\omega+k};$$
$$ C_3=-(\omega+k)(A^2-4{\Omega^{\rm F}}^2);$$
$$ A=k-\omega+2m\Omega^{\rm F}-(\omr)^2(\omega+k).$$
Let us expand Eqs. (5) near the Alfv\'en surface, i.e. near
the point $r_{\rm A}$ where $A(r_{\rm A})=0$. Denoting $x=r-r_{\rm
A}$ we get
$$
\left\{
\begin{array}{l}
\fracb{d\xi}{dx}=\fracb{1}{A'x}(C_1\xi-C_2p_*);\\
\fracb{dp_*}{dx}=\fracb{1}{A'x}(C_3\xi-C_1p_*),
\end{array}
\right.\eqno (6)
$$
where $\left.A'=\fracb{dA}{dr}\right|_{r=r_{\rm A}} $. The general
solution of these equations is
$$ \left(\xi\atop p_*\right)=\alpha_1\left(C_2\ln x\atop C_1\ln x-A'\right)+
\alpha_2\left( C_2 \atop C_1\right), \eqno (7) $$ where $\alpha_1$
and $\alpha_2$ are constants. They are not arbitrary; rather they
are fixed by the boundary conditions on $\xi$ and $p_*$ obtained
from solving general equation (5). We see that $\xi$ and $p_*$
have a logarithmic singularity at the Alfv\'en point. Later we
adhere to our expansion over $r-r_{\rm A}$, assuming $r=r_{\rm
A}$, $A=0$ whenever the quantities under consideration have no
singularities at $r=r_{\rm A}$. Let us calculate $d{\bf r}/dt$ and
$(d\e/dt)_{\rm e.m.}$ to first order in the fields of the wave.
Remembering that ${\bf E}=0$ in our reference frame and knowing
that terms with the energy $\e$ contain a small parameter, either
the ratio of the Larmor radius to the jet radius or to the
characteristic length of variations of zero order fields, we can
simplify the drift equations as follows,
$$\left(\frac{{d\bf r}}{dt}\right)_0=v_\|{\bf e}_z,$$
$$\left(\frac{{d\bf r}}{dt}\right)_1=v_\|\frac{{\bf B}_1}{B}-
                             v_\|\frac{{\bf B(B\cdot B}_1)}{B^3}+
                             \frac{[{\bf E}_1\times{\bf B}]}{B^2}\mbox{.} \eqno (8)$$
The subscript 0 denotes the zero-order term and 1 denotes the
first order term. The force of inertia is also present in the
expression of ${d\bf r}/{dt}$ (see Sivukhin, 1965). The term with
inertia force contains the electron mass and has the same order as
terms with energy and is omitted. It is clear that in the
force-free approximation the impact of the inertia force to the
trajectory is negligible. When we substitute ${\bf B}_1$ and ${\bf
E}_1$ into Eq. (8) we get
$$\left(\frac{{dr}}{dt}\right)_1=(v_\|-1)\frac{i\xi F}{\gamma}, \eqno(8*) $$
where $\gamma=\sqrt{1+(\omr)^2}$ is the Lorentz factor
corresponding to the velocity of the moving frame, $\xi$ depends
on $z,\phi,t$ by $\exp\{-i\omega t+ikz+im\phi\}$. Using the
Lorentz transformation, the phase of the wave $\Psi$ ''seen'' by
the particle is found to be $\displaystyle \Psi=-\frac F\gamma
t'+\frac F\gamma z'$. Assuming $ dz/dt=v_\| $ we get
$\displaystyle \Psi=(v_\|-1)\frac F\gamma t'$. As we see, the
coefficient for the phase coincides with the coefficient in
Eq.~(8*). This confirms our calculations because $\xi$ is the
Lagrangian displacement. The question of the trajectory of the
particle will be considered later.

Now we proceed with calculating $({d\e}/{dt})_{\rm e.m.}$:
$$ \left(\frac{d\e}{dt}\right)_{\rm e.m.}=
e{\bf E}_1\cdot\left(\frac{{d\bf r}}{dt}\right)_0+ \frac{\e
v_\perp^2}{2B}\frac{\partial B_1}{\partial t}.\eqno (9) $$ After
substitution of expressions for fields into (9) we get
$$ \left(\frac{d\e}{dt}\right)_{\rm e.m.}=\e v_\perp^2
\left(-\frac{iF}\gamma\right)\frac\xi{rS}(\omega+k)[F-\omega],\eqno
(10) $$ or, combining with (8*)
$$ \left(\frac{d\e}{dr}\right)_{\rm e.m.}=
\frac{\e
v_\perp^2}{1-v_\|}\left(\frac{(\omega+k)[F-\omega]}{rS}\right).\eqno
(11) $$ The inertia force gives the term without the coefficient
 $(\omega+k)[F-\omega]/S$. This coefficient
is large if the fast magnetosonic resonance surface $S=0$ happens
to be close to the Alfv\'en resonance $r_{\rm A}$. Our
calculations (Istomin~\& Pariev 1996) show that $S$ is indeed
small at $r=r_{\rm A}$ when the damping rate of the eigenmodes is
maximal. The maximal damping rate is still small compared to the
frequency $\omega$. Rapid damping of energy in these modes leads
to a high efficiency of particle acceleration, which depends, as
we will see later, on the smallness of $S$ ($Q\propto 1/S$ in
Eq.~(23) must be large in order to get high particle
acceleration). The inertia force is of the order of $\e/r$, which
may not give large acceleration
 because $\Delta r<<r_{\rm A}$.
As we see, $\e$ is proportional to $r$ if we do not take into
account the dependence of the right hand side on $\e$. (Note again
that we consider all values at the Alfv\'en
 resonance, $r=r_{\rm A}$.)

Therefore it makes sense to consider the motion of the particle
averaged over many periods. Taking the $r$-component of (8*), we
have
$$\frac{dx}{dt}=i\omega^*\xi(x)e^{i\omega^*t},\quad \omega^*=(v_\|-1)\frac F\gamma. \eqno (12)$$
Near the Alfv\'en resonance $\xi(x)$ has a logarithmic singularity
(Eq.~(7)); therefore, Eq.~(12) can be written as
$$\frac{dx}{dt}=\omega^*[A_1\ln x\sin(\omega^*t)+B_1\cos(\omega^*t)], \eqno(13)$$
where $\omega^*A_1$ and $\omega^*B_1$ are the dimensionless
amplitudes of the perturbation \label{ampl}(remember that $c=1$).
 Let us average Eq.~(13), expanding it in small amplitude of oscillations
of a particle in the field of the wave. We denote quantities of
zero, first, second order as $x_0$, $x_1$, $x_2$.
 $$\frac{dx_1}{dt}=\omega^*[A_1\ln x_0\sin(\omega^*t)+B_1\cos(\omega^*t)], $$
 $$x_1=-A_1\ln x_0\cos(\omega^*t)+B_1\sin(\omega^*t), $$
 $$\frac{dx_2}{dt}=\omega^*[A_1\ln|x_0-A\ln x_0\cos(\omega^*t)+B_1
\sin(\omega^*t)|\sin(\omega^*t)+B_1\cos(\omega^*t)], $$
 $$\overline{\frac{dx_2}{dt}}=\omega^*A_1B_1\frac 1{2x_0}.$$
 Now let us replace $x_0$ by ${\bar x}$ and $\displaystyle
\overline{\frac{dx_2}{dt}}$ by $\displaystyle \frac{d{\bar
x}}{dt}$ to obtain a smoothed equation of motion
 $$ \frac{d\bar x}{dt}=\omega^*\frac {A_1 B_1}{2\bar x},$$
 which has the solution
 $$ \bar x=\sqrt{\omega^*A_1 B_1 t}. \eqno(14)$$
The result for the smoothed particle motion~(14) was checked by
the direct numerical integration of Eq.~(13).

From Eqs. (11), (14) it follows that the particle gains energy drifting
along the $r$ axis. The rate of acceleration is proportional to
the transverse momentum. Because of synchrotron losses, $p_\perp$
and $J_{\perp}$ are monotonically decreasing according to
($\displaystyle\omega_B=\omega_{B0} m_{\rm e} c^2/\e$)
$$ \frac{dJ_\perp}{dt}=-\frac 43\omega_B^2\frac {r_{\rm e}}c\left(
\frac\e{m_{\rm e} c^2}\right)^3
   \left(1-\frac{v_\|^2}{c^2}\right)J_\perp, \eqno (15)$$
this  leads to $p_\|>p_\perp$. Such an anisotropic distribution
can become unstable to the excitation of electromagnetic waves
which  results in isotropization of the distribution function in
momentum space. The synchrotron losses and anisotropic
instabilities lead to the violation of the adiabatic invariant
conservation $J_\perp = \mbox{constant}$.

First of all we consider instabilities in a magnetically dominated
plasma with parameter $\beta=4\pi nT/B^2<<1$ where T is the mean
plasma energy, because the jet remains force--free when particles
gain small energies in the process of acceleration. Parail \&
Pogutse~(1986) show that instabilities may arise from the
anomalous Doppler resonance of the waves with the accelerated tail
of the distribution function. The expression they find for the
increment of instability is
\def\omk{\omega_{\bf k}}
$$
 \gamma_{\bf k}=\frac\pi 2\omk\frac{\omega_{\rm p}^2}{k^2}m_{\rm e}^2\int d{\bf p}\,
   \delta(\omk-k_{\parallel}v_{\parallel})\frac{\partial f}{\partial
p_{\parallel}}k_{\parallel}+
   \frac\pi 8\omk\frac{\omega_{\rm p}^2}{k^2}\times
$$
$$
\int d{\bf p}\,
   \delta(\omk+\omega_{B}-k_{\parallel}v_{\parallel})
\frac{k_\perp^2p_\perp^2}{\omega_{B0}^2}
   \left(k_{\parallel}\frac{\partial f}{\partial p_{\parallel}}-
\frac{m_{\rm e}\omega_{B0}^2}{p_\perp}
   \frac{\partial f}{\partial p_\perp}\right). \eqno (16)
$$
Here $\omega_{B}=\omega_{B0}(1-v^2/c^2)^{-1/2}$ and
$f=f(p_{\parallel}, p_{\perp})$ is the distribution function of
particles in drift variables $p_{\parallel}$, $p_{\perp}$. The
first term corresponds to the Cherenkov resonance and the second
corresponds to the anomalous Doppler resonance. This expression
was written with the assumption that $k_\perp p_\perp/m_{\rm
e}\omega_{B0}<1$. Parail \& Pogutse~(1986) demonstrate the
instability for the distribution function of electrons accelerated
in an electric field and call it the ``fan'' instability. This
instability is also present for all distribution functions with
small $p_\perp$ and large $p_{\parallel}$. We consider a
particular distribution function, Gaussian in $p_\perp$ and
power-law with AN index of $\alpha$ in $p_{\parallel}$ with the
minimal longitudinal momentum $p_{\rm min}$. The resulting
$\gamma_{\bf k}$ is expressed as:  $$ \gamma_{\bf k}=\frac\pi
2\frac{\omk}{\omega_{B0}}\frac{\omega_{\rm p}^2}{k^2}
\frac{\alpha-1}{p_{\rm min}}m_{\rm e} \left(\frac{p_{\rm
min}k_{\parallel}}{\omega_{B0}m_{\rm e}}\right)^\alpha
k_{\parallel}\left[-\alpha\left(\frac{\omega_{B0}}{\omk}\right)^{\alpha+1}
\gamma_1^{2-\alpha}+
\frac12\frac{k_\perp^2}{k_{\parallel}^2}\gamma_2^3\right], \eqno
(17)
$$
where $\gamma_1$, $\gamma_2$ are the $\gamma$-factors of the
particles in Cherenkov resonance and anomalous Doppler resonance
correspondingly. If $\gamma_1\approx\gamma_2\approx1$ (the
particles in resonance are non-relativistic) we see that
instability arises when $\displaystyle
2\alpha\left(\frac{\omega_{B0}}{\omk}\right)^{\alpha+1}<
  \frac{k_\perp^2}{k_{\parallel}^2}<\frac{m_{\rm e}^2\omega_{B0}^2}
{k_{\parallel}^2p_\perp^2}$, or $\displaystyle
p_\perp^2<\frac{m_{\rm e}^2}{2\alpha}\frac{\omega_{B0}^2}
{k_{\parallel}^2}
\left(\frac{\omega_{B0}}{\omk}\right)^{\alpha+1}$ i.e. if
$p_\perp$ is small enough. Quasilinear analysis of the
influence of the excited waves on the
 distribution function shows (Parail \& Pogutse, 1986)
that the developing instability leads to the isotropization of the
distribution function. This may easily be seen from the resonance
condition $\hbar\omk=n\hbar\omega_{B}+\hbar
k_{\parallel}v_{\parallel}$ ($n$ is an integer), which may be
treated as the energy conservation law for one interaction of a
particle with the wave. Here $\hbar k_{\parallel} v_{\parallel}
\approx-[(m_{\rm e} v_{\parallel}-\hbar k_{\parallel})^2- m_{\rm
e} ^2v_{\parallel}^2]/(2m_{\rm e})$ is the decrement of the
longitudinal energy, which is positive in the case of anomalous
Doppler resonance ($n<0$). So the longitudinal energy decreases,
and the transverse energy increases ($n<0$). In the case of
$\omk<<\omega_{B}$ the energy transmitted to the waves is small.
Due to this instability the distribution function of electrons
becomes isotropic with deviations from isotropy of the order of
$\displaystyle \simeq \frac{\omega_{B0}}{\omega_{\rm p}^2\tau}$,
where $\tau$ is the smallest characteristic time of non-stationary
processes increasing the anisotropy of the distribution function,
and the plasma frequency is $\omega_{\rm p}^2= 4\pi n e^2/m_{\rm
e}$. Given typical values of $B_0 \approx 10^{-2}\,\mbox{G}$,
$n\approx 0.1\,\mbox{cm}^{-3}$ we have $\omega_{B0}/ \omega_{\rm
p}^2\approx 10^{-3}\,\mbox{sec}$. This time is many orders of
magnitude smaller than any possible $\tau$ either due to
synchrotron losses $\tau_{\rm s}$ or $\approx 1/\omega$, or
$\approx 1/\Omega^{\rm F}$. Therefore, the distribution of
accelerated particles must be highly isotropic in momentum space.

Although the case with $\beta>1$ (strong accelerated particles) is
apparently contradictory with our assumption of a force-free jet,
the process of acceleration takes place in the close vicinity of
the Alfv\'en resonance and does not affect the overall structure
of the fields in the whole jet. As we show later the process of
acceleration  leads to the power-law spectrum of particles  with
high characteristic energy (see Eq.(22)) and the minimum
energy a few orders of magnitude lower. So the case with $\beta>1$
can occur.

In the case of $\beta>1$ there exist many types of instabilities,
some of which are hydrodynamic.
 We will treat plasma instabilities according to Mikhaylovskii~(1975),
who showed that in the case $p_\|>p_\perp$, perturbations with
$k_{\parallel}=0$ have maximal growth rates. The frequencies of
these perturbations with
\def\oureps{\varepsilon_{33}}
${\bf k}\perp{\bf B}$ are the solutions of the dispersion relation
$\oureps-N^2=0$, where $\oureps$ is the component of the
dielectric tensor along the magnetic field, $N=ck/\omega$. The
dispersion relation of the small oscillations has the form
\def\pa{\partial}
\def\om{\omega}
\def\omb{{\omega_B}}
\def\der#1#2{\frac{{\pa #1}}{{\pa #2}}}
$$\frac{c^2k^2}{4\pi e^2}=\frac{\om^2}{4\pi e^2}+
  \int\,d{\bf p}\left\{ v_{\parallel}
\left[\frac{\pa f}{\pa p_{\parallel}}-\left(1-J_0^2
(kv_{\perp}/\omega_B)\right)
  \frac{v_{\parallel}}{v_\perp}\frac{\pa f}{\pa p_\perp}\right]\right.$$
$$ \left.+2\om^2\sum_{n=1}^\infty\frac{\frac{v_{\parallel}^2}{v_\perp}
  \frac{\pa f}{\pa p_\perp}}{\om^2-n^2\om_B^2}J_n^2
(kv_{\perp}/\omega_B)\right\}.\eqno (18)
$$
The anisotropic instabilities result in the excitation of
potential waves which change the direction of the particle
momentum $\bf{p}$ but do not change the energy $\e$ of the
particle. Indeed, momentum conservation in the process of emitting
or absorbing a wave reads
$$
\delta{\bf p} = \hbar{\bf k} =
\hbar\omega\left(\frac{N}{c}\frac{{\bf k}}{k}\right),
$$
where $N$ is the wave refractive index. For the potential waves,
$N>>1$. If $\delta{\bf p}$ is not orthogonal to ${\bf p}$, then
$\delta\e \approx|\delta{\bf p}|\cdot c = \hbar\omega\cdot N
>>\hbar\omega$. This is in contradiction to the energy
conservation $\delta\e = \hbar\omega$. So, $\delta{\bf p}\perp{\bf
p}$ with the accuracy $1/N << 1$.

We see that the effect of anisotropic instability is to change the
particle distribution function in the direction of momentum, and not
in energy.
Because the instability is rapid, the quasi-stationary
distribution function is determined from the condition
$\gamma_{\bf k} = 0$. This condition is the equation which should
be used instead of Eq.~(15) describing the change of
$J_\perp$ only due to synchrotron losses. The condition
$\gamma_{\bf k} = 0$ is the integral equation for the particle
distribution function $f(p_\parallel , p_\perp)$, which follows
from Eq.~(18). It is difficult to solve this integral
equation without knowledge about the shape of
$f(p_{\parallel},p_{\perp})$. We estimate the integral in the
dispersion relation~(18) assuming the cold distribution function
$f$, although in reality it is smooth in momentum space. Also, we
assume that $k v_\perp$ is of the order of $\omb$ and estimate
expression~(18) taking into account only the first terms of
The expansion in Bessel functions over its arguments. Then, the
threshold of stability ($\gamma_{\bf k}=0$) when particles have
equal $p_\perp$ and $p_\|$ in the ultrarelativistic case is found
to be
$$ \frac{p_{\parallel}^2}{p_\perp^2}=\frac{4-\beta+
\sqrt{\beta(17\beta-8)}}{4(\beta-1)}, \eqno (19)$$ where
$\beta=4\pi nm_{\rm e} c^2\gamma/B^2$.
 We denote the threshold of stability $p_{\parallel}/p_\perp=\alpha$,
$\alpha>1$. Expression~(19) is not quite valid for very large
$\beta(>8)$ but large $\beta$ are not achieved here. $\gamma_{\bf
k}>0$ arises when $p_{\parallel}>\alpha p_{\perp}$. The
instability is fast because its growth rate is proportional to the
cyclotron frequency; still the growth rate is much less than the
cyclotron frequency. We will assume that the process of
acceleration and the fast instability
 considered above will form a nonequilibrium distribution function with
anisotropy of order unity. However, after particles leave the
acceleration region ($\beta>1$), various instabilities such as the
``fan" instability considered in this section lead to fast full
isotropization of the distribution function.

Now we understand that ${dJ_\perp}/{dt}\ne 0$ not only  because of
the synchrotron losses but also because of the anisotropic plasma
instability. We use the relation
$p_{\parallel}/p_\perp\simeq\alpha$ as an approximate closure
equation for the system of two equations~(1), and express the
longitudinal and perpendicular components of the particle momentum
through the energy by the relations
$$
p_{\perp}^2=\frac{\e^2-m_{\rm e}^2 c^4}{(1+\alpha^2)c^2}; \quad
p_{\parallel}^2=\frac{\alpha^2(\e^2-m_{\rm
e}^2c^4)}{(1+\alpha^2)c^2}.
$$
Since the dependence of $\alpha$ on $\beta$ (equation~(19)) is
weak for $\beta>2$, the value of $\alpha$ is approximated as a
numerical constant independent of $r$ in further expressions for
particles trajectories and particles distribution function. For
the derivative of the energy we write equation~(11) adding the
synchrotron losses.
$$ \frac{d\e}{dt}=\frac{\e v_\perp^2}{1-v_\|}\frac{(\omega+k)[F-\omega]}{rS}
\frac{d{\bar x}}{dt}-\frac23 \frac{p_\perp^2}{m_{\rm
e}}\om_{B0}^2r_{\rm e} .\eqno (20) $$ Substituting $d{\bar x}/dt$
from (14) we obtain
$$ \frac{d\e}{dt}=\frac\e{1+\alpha^2-\alpha\sqrt{1+\alpha^2}}
\frac{(\omega+k)[F-\omega]}{rS}\left(\frac{\om^*A_1
B_1}{4t}\right)^{1/2}
-\frac23\frac{\e^2}{1+\alpha^2}\om_{B0}^2\frac{r_{\rm e}}{m_{\rm
e}} .\eqno (21) $$
 Let us introduce some notations for convenience:
$$ Q=\frac1{1+\alpha^2-\alpha\sqrt{1+\alpha^2}}
     \frac{(\omega+k)[F-\omega]}{S};  $$
$$ \e_1=\frac34\frac{m_{\rm e}}{1+2\alpha^2-2\alpha\sqrt{1+\alpha^2}}
\left(\frac{(\omega+k)[F-\omega]}S\right)^2 \frac{{\om^*}^2A_1
B_1}{\om_{B0}^2r_{\rm e}r^2\om^*}. \eqno (22) $$ Here
$(\omega+k)[F-\omega]/S $ is fixed for given wavenumbers of the
perturbation, $m$ and $k$. It is large near the fast magnetosonic
resonance surface $S=0$. For an AGN jet where
 $B\approx10^{-2}\,\mbox{G}$ (Begelman at al., 1984),
$r\approx 1\,\mbox{pc}$, the coefficient $1/(\om_{B0}^2r_{\rm
e}r^2\om^*) \approx 10^4$, which is a large number.
 The value of ${\om^*}^2A_1 B_1$
is the dimensionless amplitude of the perturbation squared as we
noted after Eq.~(13). Now taking into account our notations,
the solution for $\e$ has the form
$$ \e=\left(\frac1{\e_1}\left(Q\frac xr-1\right)+\left(\frac1{\e_1}+
\frac1{\e_0}\right)\exp\left\{-Q\frac xr\right\} \right)^{-1}.
\eqno (23) $$ Here $\e_0=\e(x=0) $. Let us denote $Qx/r=x'$. If
$x'$ is of order unity, then $\e \simeq \e_0$ and particle
acceleration is not effective. We assume that $x'$ can
 be large because the quantity $Q$ is large in the case of maximal damping
rate of eigenmodes. If $x'$ is large, then $\e \gg \e_0$ and the
particle acceleration is effective.
 It is seen that first $\e$  increases exponentially and then
 decreases as $1/x'$.
 We consider the asymptotic behaviour of $\e$ in the case of different
 $x'$ and $\e_0$.
 In the case of $\e_0<\e_1$
 $$ \e=\left\{\begin{array}{rrr} \label{asym}
               \e_1/x',&\quad&x'e^{x'}>\e_1/\e_0\\
               \e_0e^{x'},&\quad&x'e^{x'}<\e_1/\e_0\\
                \end{array}\right. \mbox{.}  \eqno (24) $$
 In the case of $\e_0>\e_1$
 $$ \e=\left\{\begin{array}{rrr}
               \e_1/x',&\quad&x'>1\\
               2\e_1/{x'}^2,&\quad&x'<1\\
                \end{array}\right.  \mbox{.} $$
We ignore the case when $\e_0>\e_1$ since the typical initial
energy is much less then $\e_1$. Knowing the particle
trajectories (Eqs.~(14) and~(23)), we can find the
distribution function $F(\e,x)$ of accelerated particles in the
phase space of $(\e,x)$ using the fact that the trajectories of
the particles are the integrals of the collisionless kinetic equation.

\section{Formation of the spectrum of accelerated particles}

Our goal is to calculate the distribution function averaged over
$x$  knowing the distribution function at $x=0$. We will consider
the stationary case, when $\pa f/\pa t=0$. Let the distribution
function be given at  point $x'_0$ as a function of
$\e'=\e/\e_1$. Then the number of particles
$$dN=F_0(\e'_0,x'_0)d\e'_0dx'_0=F_0\Bigl(\e'_0(\e',x'),x'_0(\e',x')\Bigr)
 \frac{D(\e'_0,x'_0)}{D(\e',x')}d\e'dx'. $$
Below we scale $x$ with $r/Q$ and $\e$ with $\e_1$ and omit primes
in the dimensionless $x'$ and $\e'$. Thus if we know the
trajectory $\e_0(\e,x),x_0(\e,x)$ we may obtain the distribution
function at the point $\e,x$. In our case the Jacobian
$D(\e_0,x_0)/D(\e,x)$ is not equal to unity because the variables
$\e,x$ are not canonical. So,
$$ F(\e,x)=F_0\Bigl(\e _0(\e ,x ),x _0(\e ,x )\Bigr)
 \frac{D(\e _0,x _0)}{D(\e ,x )}. \eqno (25) $$
If we repeat the derivation of Eq.~(23) with an arbitrary
reference point $x_0\neq 0$ ($\e(x_0)=\e_0$) we obtain
$$\e=\{x-1+\left(\frac{1}{\e(x_0)}-x_0+1\right)\exp(-x+x_0)\}^{-1}. \eqno(26) $$
Substituting $\e_0$ expressed from (26) into (25) we obtain
\def\der#1#2{\frac{{\pa #1}}{{\pa #2}}}
$$ F(\e,x)=F_0\left(\left\{-1+x_0+\left(\frac1\e-x+1\right)\exp(x-x_0)\right\}^{-1},x_0\right)
   \left(\der{\e_0}\e\der{x_0}x-\der{\e_0}x\der{x_0}\e\right)= \eqno(27)$$
$$ F_0\left(\left\{-1+x_0+\left(\frac1\e-x+1\right)\exp(x-x_0)\right\}^{-1},x_0\right)
   \frac{\e_0^2}{\e^2}\exp(x-x_0)\frac x{x_0}. $$
We know that initial fast particles are injected near the point
$x=0$ \label{init}. Therefore in the last formula we let $x_0=0$
everywhere except in the denominator. Physically
$F_0(\e_0,0)$ is the particle density at zero point and $x_0$ is
the distance below which  our formulae, for instance~(14), are not valid.
Because of the small imaginary part of $\omega$ which
is an attenuation increment of the wave~(Istomin \& Pariev, 1996),
the divergence $\ln x$ at the resonance is cut at the point
$$ |r_0-r_{\rm A}|\approx\left|\frac{A(r=r_{\rm A})}{dA/dr(r=r_{\rm A})}\right|
\approx \left|\mbox{Im}\,\omega \frac{r_{\rm
A}}{\mbox{Re}\,\omega}\right| ,\eqno (28)$$ where
$\mbox{Im}\,\omega \ll \mbox{Re}\,\omega$. Here $A(r=r_{\rm A})$
is not equal to zero as we assumed earlier, but determined by the
small value of the imaginary part of $\omega$. Thus, $x_0$ is the
width of the resonance $ x_0=Q(r_0-r_{\rm A})/r_{\rm A}$.
Eventually, Eq.~(27) is
$$ F(\e,x)= F_0\left(\left\{-1+\left(\frac1\e-x+1\right)\exp(x)\right\}^{-1}
,x_0\right)
   \frac{\e_0^2}{\e^2}\exp(x)\frac x{x_0}.  \eqno (29) $$
Because the width of the resonance $x_0$ and the width of the
acceleration region $\Delta r$ \label{width}, which is  some
function of $x_0$, are small compared with the jet radius we
average the distribution function over the  jet radius near the
resonance surface
$$ \bar F(\e)=\frac1{\Delta x}\int\limits_0^\infty F(\e,x)\,dx,\quad \Delta x=
\frac Q{r_{\rm A}}\Delta r. \eqno (30) $$ Denoting $N(\e)=\bar
F(\e)\Delta x$ we obtain
$$ N(\e)=\int\limits_0^\infty F_0
\left(\left\{-1+\left(\frac1\e-x+1\right)\exp(x)\right\}^{-1},
x_0\right) \frac{\e_0^2}{\e^2}\exp(x)\frac x{x_0}\,dx. \eqno(31)
$$ Let us change the variable of integration to $\e_0$. The dependence
of $\e_0$ versus $x$ under fixed $\e$ is presented in Fig.~1.
\begin{figure}
\centering
\includegraphics[width=\textwidth]{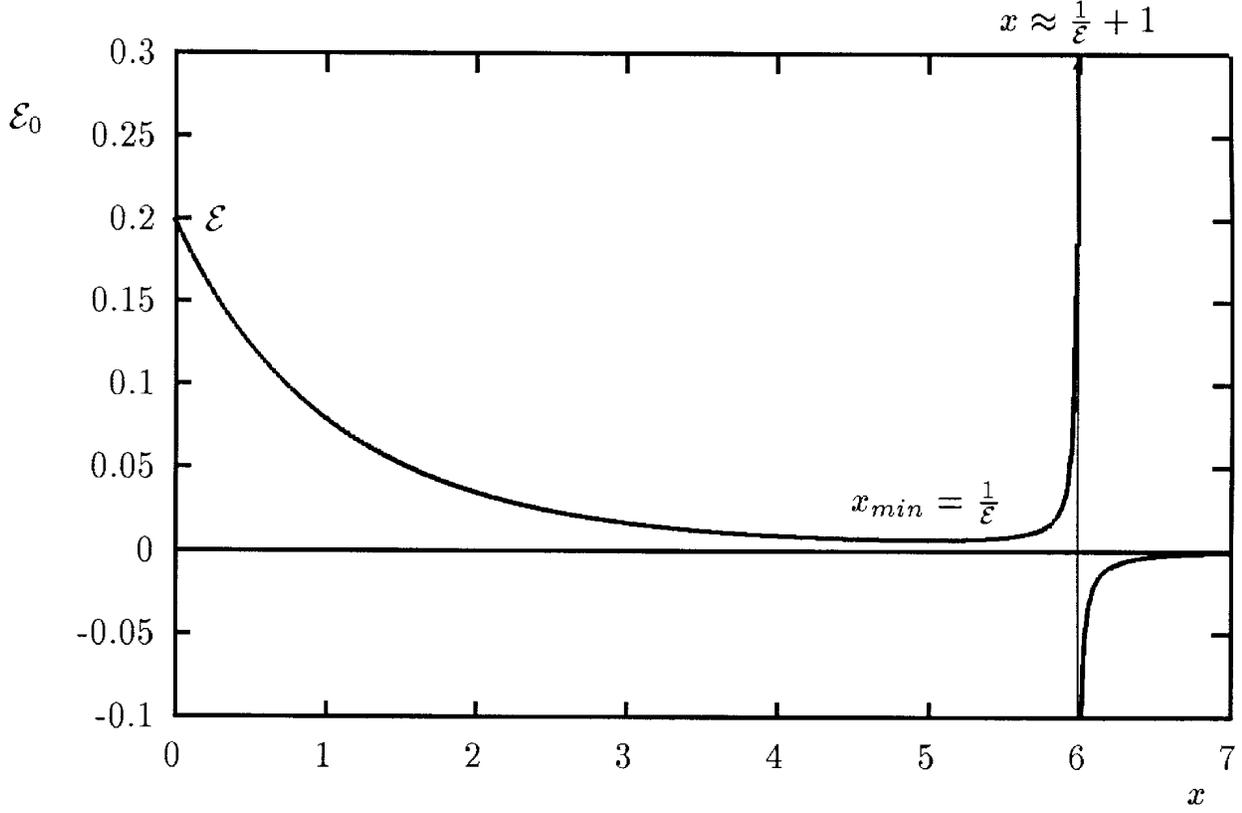}
\caption{The dependence of $\e_0$ on $x$ (the energy $\e$ and
the distance $x$ are in units of $\e_1$ and $r/Q$ (22)
correspondingly)  } \label{fig1}
\end{figure}

The value of $\e_{\rm min}$ is
$$ \e_{\rm min}=1/(-1+\exp(1/\e)). \eqno (32)$$
Because the function $x=x(\e_0)$ has two branches, the
equation~(31) is transformed into
\def\e{{\cal E}}
$$ N(\e)=\frac1{\e^2x_0}\int\limits_{\e_{\rm min}}^\e F_0(\e_0,x_0)
\frac x{1/\e-x}\,d\e_0+$$
$$ \frac1{\e^2x_0}\int\limits_{\e_{\rm min}}^\infty F_0(\e_0,x_0)
\frac x{1/\e-x}\,d\e_0,\eqno(33)$$ where $x$ is taken on the first
branch of $x(\e_0)$ in the first integral and on the second branch
in the second integral. Let us make some natural assumptions. The
initial distribution function is cut off at some energy of
${T_0}$, provided that $T_0 \ll 1$ (bearing in mind that $T_0$ as
well as $\e$ are in the units of energy $\e_1$~(22)). Let us
calculate $N(\e)$ in the region, where $T_0\ll\e\ll 1$, because it
is seen from Eqs.~(32) and (33) that in the case $\e\sim1$
the integral is small.
 Let us evaluate the contribution of the singularity $x=1/\e$ in integrals~(33)
$$ \e_0-\e_{\rm min}=\frac{d^2\e_0}{dx^2}\left(x-\frac1{\e}\right)^2\mbox{.} $$
The integral near the singularity is
$$ \frac1{\e^2} \int\limits_{\e_{\rm min}} F_0(\e_{\rm min},x_0)\frac{\frac1{\e}\sqrt{
\frac{d^2\e_0}{dx^2}}}{x_0\sqrt{\e_0-\e_{min}}}\,d\e_0\sim\sqrt{\e_0-\e_{\rm
min}}. $$ Thus, its contribution is small. Because the main
contribution to Eq.~(26) is when $x$ is far from $1/\e$, we
may use asymptotic values of $x$ (see Eq.~(24)): $
x\approx\ln(\e/\e_0) $  on the first branch; $ x\approx1/\e+1 $ on
the second branch and obtain
$$ N(\e)=\frac1{\e^2x_0}\int\limits_{\e_{\rm min}}^\e F_0(\e_0)
\frac{\ln(\e/\e_0)}{1/\e-\ln(\e_0/\e)}\,d\e_0
+\frac1{\e^2x_0}\int\limits_{\e_{\rm min}}^\infty F_0(\e_0)\frac
1\e\,d\e_0.\eqno(34)$$ Let us consider only $\e_{\rm min}\ll T_0$
i.e. $\e\ll-1/\ln T_0$ because in the reverse case the integral is
small. Now we can replace $\e_{\rm min}$ by~$0$ and also
$1/\e\gg\ln(\e/\e_0)$. Eq.~(34) becomes
$$ N(\e)=\frac1{\e x_0}\int\limits_0^\e F_0(\e_0)(-\ln \e_0)\,d\e_0+\frac n{\e^3 x_0}, \eqno(35) $$
where $n=\int_0^\infty F_0(\e_0)\,d\e_0$. Rewriting
Eq.~(35) as averaging $<\ldots >$ over the initial
distribution function $F_0$
$$ N(\e)=\frac n{\e x_0}<\! -\ln \e_0 \! >+ \frac n{\e^3 x_0}, \eqno (36) $$
we note that the first term is approximately equal to
$$ \frac n{\e x_0} (-\ln T_0), $$
and is small compared to the second term because the inequality
$1/\e^2 \gg -\ln T_0$ is the consequence of inequality $1/\e \gg
-\ln T_0$. Therefore $N(\e)\propto 1/\e^3 $, and we obtain a power-law 
distribution with the index~$-3$. The validity region of this
formula $T_0<\e<-1/\ln T_0$ is quite large.

We performed the integration in Eq.~(33) numerically, with
Gaussian initial distributions at different temperatures. The results
are presented in Figs.~2 and~3.

\begin{figure}
\centering
\includegraphics[width=\textwidth]{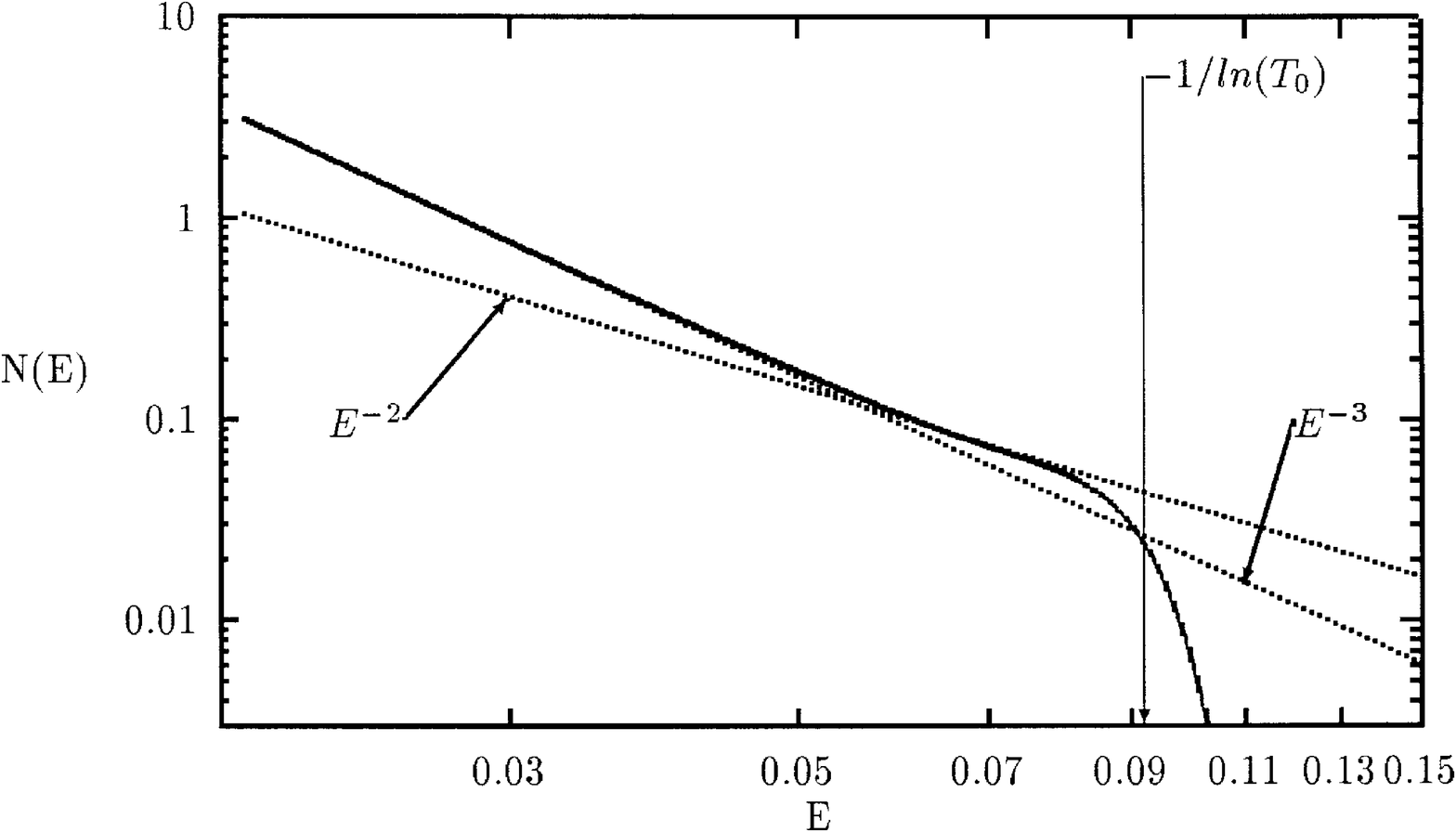}
\caption{Distribution function calculated with a Gaussian initial
distribution with temperature $2\cdot10^{-5}\e_1$ (the energy $\e$
is in units of $\e_1$ (22)) } \label{fig2}
\end{figure}

\begin{figure}
\centering
\includegraphics[width=\textwidth]{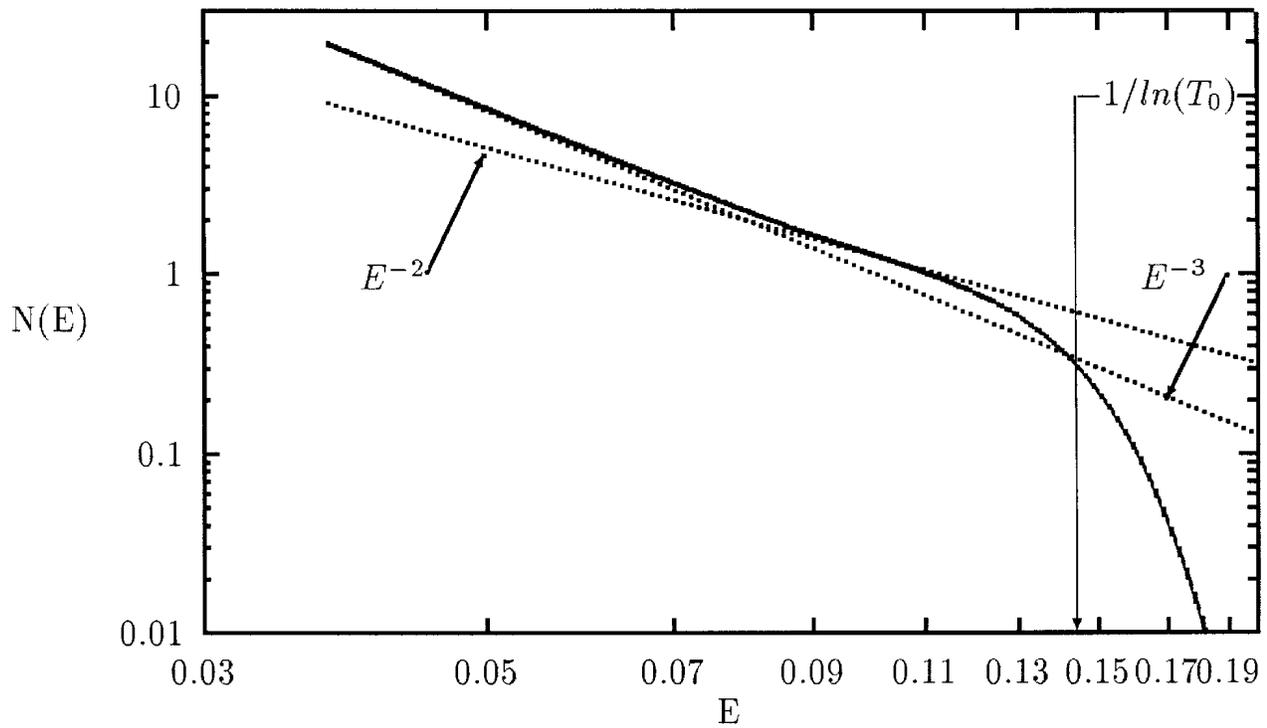}
\caption{Distribution function calculated with Gaussian initial
distribution with temperature $10^{-3}$ (the energy $\e$ is in 
units of $\e_1$ (22)) } \label{fig3}
\end{figure}

The value of the power-law index is $-3$ for small energies and
rises slightly before the cutoff of the distribution function.
However, the $-3$ power-law index is not applicable all the way
down to $\e=0$. The natural boundary is the width of the
acceleration region, $\Delta r$ (see Eq.~(33)). Because the
main contribution to the integral~(28) is from the region near the
point $x=1/\e+1$, the lower limit for power law $1/\e^3$ is
$\e_{\rm l}=1/\Delta x$. We also found the mean power law index
for the distribution of particles between energies $\e_{\rm l}$
and $-1/\ln T_0$. The results are presented in Fig.~4.

\begin{figure}
\centering
\includegraphics[width=\textwidth]{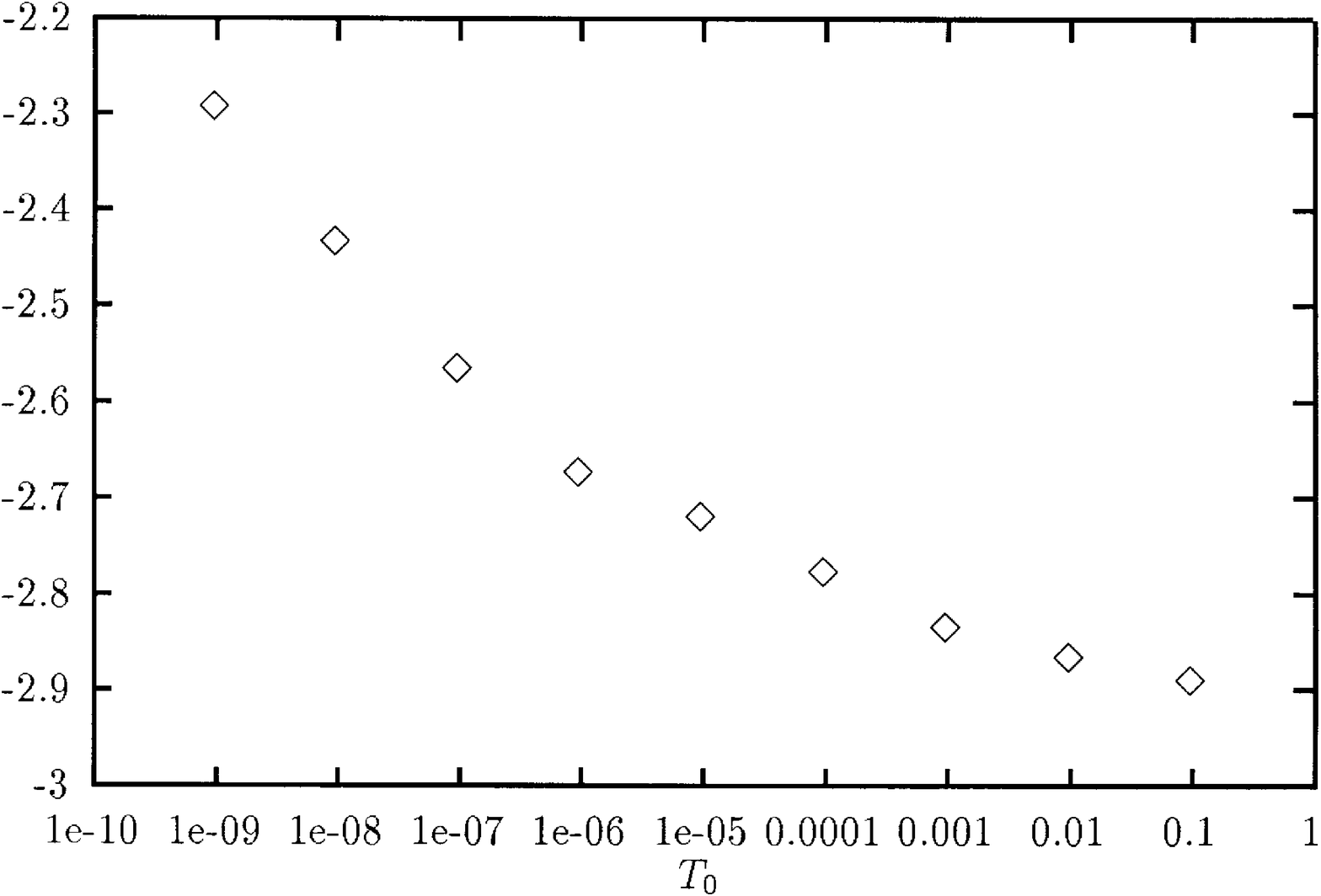}
\caption{Averaged power law index (the initial particle energy
$T_0$ is in units of $\e_1$ (22)) } \label{fig4}
\end{figure}

Recall that in the acceleration region the distribution function
of the accelerated particles has an anisotropy of order unity, but
the distribution of particles escaping the acceleration region
becomes isotropic in momentum space very quickly.

One remaining question in the formation of the spectrum of
energetic particles is the origin \label{origin}of the initial
particles, since to first order in the perturbed fields there is
no acceleration of the cold plasma in Eqs.~(1). We have
already mentioned that the particles to be accelerated originate
in the region close to the Alfv\'en resonance. It can be seen from
Eqs.~(1) that to second order in the perturbed fields,
acceleration always takes place and the power acquired corresponds
to the term ${\bf j\cdot E}$ in the energy conservation equation
for the electromagnetic field. The damping of the wave occurs
due to the heating of the plasma by current ${\bf j}_1$ which is
described by second-order terms. If all of the wave energy is
transmitted to the particles, the condition $\beta>1$ would imply
that the energy density of the wave is greater than the energy
density of the stationary field. Let us assume this at least on
the surface of the resonance where the field is maximal. Then the
energy acquired by one particle can be evaluated as $T_0\approx
B_0^2/8 \pi n$. Given typical values $B_0\approx 10^{-2}\,
\mbox{G}$ and $n\approx 0.1\, \mbox{cm}^{-3}$ we obtain
$T_0\approx 10\,\mbox{MeV}$.

\section{Summary}

In the present work we considered possible acceleration of
electrons and positrons inside relativistic rotating
electron-positron force--free cylindrical jets with spiral
magnetic fields. It is very plausible that inner parsec--scale
jets in active galaxies can be described in the frame of this
physical model.
 The observations of jets in 3C273, M87, 3C279, 3C84 and PKS0528+134 show
that extragalactic jets are likely to be electron-positron (
Morrison \& Sadun, 1992; Reynolds et al., 1996; Wardle et al.,
1998 ). It also seems that such jets are not in steady state and
can contain waves excited by environmental processes, of which the
most powerful would be the variability of the accretion rate onto
the central black hole and accretion disc instabilities
(magnetorotational instability). Both strongly perturb the
magnetic field at the base of the jet.  We considered the
behaviour of such excitations inside cylindrical force--free jets
embedded in a dense medium in our previous studies (Istomin \&
Pariev, 1994; Istomin \& Pariev, 1996). We estimate that the
density of the gas commonly found in the nuclei of galaxies is
large enough that it should be taken into account when considering
the evolution of perturbations inside the force--free magnetized
jet. We found that in the wide range of parameters determining the
equilibrium structure of electromagnetic fields and perturbations
inside the jet there exist resonant surfaces on which the phase
speed of eigenmodes is equal to the local Alfv\'en velocity. Those
eigenmodes for which there is an Alfv\'en resonance have a small
damping rate, whilst those having no Alfv\'en resonances are
neutrally stable. This shows that the existence of an Alfv\'en
resonance leads to losses of energy of corresponding eigenmodes.

Indeed, the amplitude of the wave is large in the vicinity of the
Alfv\'en resonant surface and decreases away from the Alfv\'en
resonant surface. Both electrons and positrons are subject to
drifting out of this region of strong electric field of the wave
due to the non-uniformity of the electric field. When moving to a
region of weak electric field, particles gain energy. This energy
is taken from the energy of the wave, which is an eigenmode in the
jet, and the eigenmode decays. In this work we propose a novel
mechanism for acceleration of particles {\it in situ} in strong
electromagnetic waves of varying amplitude inside the jet. Unlike
the acceleration on shock fronts proposed to operate inside bright
knots (e.g., Blandford~\& Eichler 1987), the acceleration by
electromagnetic waves is regular, i.e. it does not require
turbulence or inhomogeneities to be present in the jet. Also, the
net gain of energy by particles in our mechanism occurs slowly on
a time scale greatly exceeding the Larmour orbital time and also
exceeding the period of the wave. In this aspect, the regular
electromagnetic acceleration by a nonuniform wave differs from
acceleration in magnetic reconnection layers (Romanova~\& Lovelace
1992; Blackman 1996), where strong departures from ideal MHD occur
and a particle gains its total energy during one passage through
the region of strong electric field. Our acceleration mechanism
does not rely upon violation of ${\bf E}\cdot {\bf B}=0$ in any
region inside the jet as the gain of energy during the drift is a
result of a kinetic treatment of plasma possible only when one
takes into account a finite Larmour radius.

We found how particles are accelerated in the strong fields near
the Alfv\'en surface. We considered the process of particle
acceleration using equations of motion in drift approximation with
addition of synchrotron losses and isotropization of particle
distribution by plasma instabilities. Such instabilities are
excited by accelerated particles if the distribution function is
anisotropic in momentum space, and act to conceal any anisotropy of
the distribution of particles. Acceleration process and
synchrotron losses taken together form a power law energy spectrum
of ultra-relativistic electrons and positrons with index between 2
and 3 depending upon the initial energy of the injected particles.
This is consistent with the typical spectral indices of radio
emission observed in parsec--scale jets.
 The power law spectrum extends up to
 the energy $\e_{\rm max}$, where a sharp cutoff occurs (see Figs.~2,3).
 The magnitude of $\e_{\rm max}$ depends on the initial particle temperature
 $T_0$ as well as on the characteristic acceleration energy $\e_1$~(26):
 $\e_{\rm max}\approx \e_1/\ln(\e_1/T_0)$. The quantity $T_0$ evaluated
 from the equipartition condition in the acceleration region is equal to
 $10\,\mbox{MeV}$  for the magnetic field $10^{-2}\,\mbox{G}$
 and density $0.1\,\mbox{cm}^{-3}$ correspondingly. The quantity $\e_1$
is, according to~(26), of the order of $10^4(\delta B/B_0)^2\,
 \mbox{MeV}$ for the value of $B_0=10^{-2}\,\mbox{G}$, where $\delta B/B_0$ is the
 dimensionless amplitude of the perturbation. For large perturbations
 ($\delta B\approx B_0$) $\e_1\sim 10^4\,\mbox{MeV}$.
 Thus particles accelerated near Alfv\'en resonance are in the energy range
 of $10\,\mbox{MeV}<\e<10^3\,\mbox{MeV}$. These particles emit synchrotron
radiation in the range of frequencies from approximately
$100\,\mbox{MHz}$ to $1000\,\mbox{GHz}$ which covers the
frequencies of modern radio observations. We present results of
the calculations of synchrotron emission and polarization produced
by the accelerated particles with power law distribution over
energy in a separate article (Pariev et al. 2002).

\begin{acknowledgements}
We thank R.I. Selkowitz for checking the English in the final version
of the manuscript. This work was done under support of the Russian
Foundation for Fundamental Research (grant number 02-02-16762). VP
acknowledges partial support from DOE grant DE-FG02-00ER54600.
\end{acknowledgements}

\newpage

\begin{center}
{\Large References}
\end{center}
\parindent0pt

Alfv\'en~H., F\"althammar~C.-G., 1963, Cosmical Electrodynamics,
Fundamental Principles. Second edition, Clarendon Press, Oxford,
p.28

Appl~S., Camenzind~M., 1992, A\&A, 256, 354

Begelman~M.C., Blanford~R.D., Rees~M.J. 1984, Rev. Mod. Phys., 56,
255

Beskin~V.S., Istomin~Ya.N., Pariev~V.I., 1992, AZh, 69, 1258

Bisnovatyi--Kogan~G.S., Lovelace~R.V.E., 1995, A\&{}A, 296, L17

Blackman~E.G., 1996, ApJ, 456, L87

Blandford~R.D., Eichler~D., 1987, Physics Reports, 154, 1

Blanford~R.D., Pringle~J.E., 1976, MNRAS, 176, 443

Frolov~V.P., Novikov~I.D., 1998, Black hole physics: basic
concepts and new developments. Kluwer Academic, Dordrecht

Hirotani~K., Okamoto~I., 1998, ApJ, 497, 563

Hirotani~K., Iguchi~S., Kimura~M., Wajima~K., 2000, ApJ, 545, 100

Istomin~Ya.N., Pariev~V.I., 1994, MNRAS, 267, 629

Istomin~Ya.N., Pariev~V.I., 1996, MNRAS, 281, 1

Li~L.-X., 2000, ApJ, 531, L111

Lyubarskii~Yu.E., 1999, MNRAS, 308, 1006

Mikhaylovskii~A.B., 1975, in Reviews of Plasma Physics, ed. by
Leontovich M.A., Consultants Bureau, New York, Vol.~6, p.~77

Morozov~A.I., Solov'ev~L.S., 1966, in Reviews of Plasma Physics,
ed. by Leontovich M.A., Consultants Bureau, New York, Vol.~2,
p.~201

Morrison~P., Sadun~A., 1992, MNRAS, 254, 488.

Mushotzky~R.F., Done~C., Pound~K., 1993, Ann. Rev. Astron.
Astrophys., 717

Parail~V.V., Pogutse~O.P., 1986, in Reviews of Plasma Physics, ed.
by Leontovich M.A., Consultants Bureau, New York, Vol.~11, p.~1

Pariev~V.I., Istomin~Ya.N., Beresnyak~A.R., 2002, A\&A, accepted

Rees~M.J., 1984, Ann. Rev. Astron. Astrophys. ,471

Reynolds~C.S., Fabian~A.C., Celotti~A., Rees~M.J., 1996, MNRAS,
283, 873.

Romanova~M.M., Lovelace~R.V.E., 1992, A\&A, 262, 26

Sikora~M., Madejski~G., 2000, ApJ, 534, 109

Sivukhin~D.V., 1965, in Reviews of Plasma Physics, ed. by
Leontovich M.A., Consultants Bureau, New York, Vol.1, p.1

Torricelli-Ciamponi~G., Petrini~P., 1990, ApJ., 361, 32

Wardle~J.F.C., Homan~D.C., Ojha~R., Roberts~D.H., 1998, Nature,
395, 457

Witzel~A., Wagner~S., Wegner~R., Steffen~W., Kirchbaum~T., 1993,
in Davis R.J. \& Booth R.S., eds Sub-arcsecond Radio Astronomy.
Cambridge Univ. Press, Cambridge, p. 159

\end{document}